\documentclass[twocolumn,a4paper]{revtex4}
\usepackage{amsfonts,amsmath,amssymb}
\usepackage{ragged2e}
\begin{document}

\newcommand{\Supertwistor}{\Cset \mathrm{P}^{3|4}}
\newcommand{\Twistorspace}{\Cset \mathrm{P}^{3}}
\newcommand{\half}{\frac{1}{2}}
\newcommand{\diff}{\mathrm{d}}
\newcommand{\ra}{\rightarrow}
\newcommand{\Zset}{{\mathbb Z}}
\newcommand{\Cset}{{\,\,{{{^{_{\pmb{\mid}}}}\kern-.47em{\mathrm C}}}}}
\newcommand{\Rset}{{\mathrm{I}\!\mathrm{R}}}
\newcommand{\gra}{\alpha}
\newcommand{\grl}{\lambda}
\newcommand{\gre}{\epsilon}
\newcommand{\zb}{{\bar{z}}}
\newcommand{\mn}{{\mu\nu}}
\newcommand{\Acal}{{\mathcal A}}
\newcommand{\Rcal}{{\mathcal R}}
\newcommand{\Dcal}{{\mathcal D}}
\newcommand{\Mcal}{{\mathcal M}}
\newcommand{\Ncal}{{\mathcal N}}
\newcommand{\Kcal}{{\mathcal K}}
\newcommand{\Lcal}{{\mathcal L}}
\newcommand{\Scal}{{\mathcal S}}
\newcommand{\Wcal}{{\mathcal W}}
\newcommand{\Bcal}{\mathcal{B}}
\newcommand{\Ccal}{\mathcal{C}}
\newcommand{\Vcal}{\mathcal{V}}
\newcommand{\Ocal}{\mathcal{O}}
\newcommand{\Zcal}{\mathcal{Z}}
\newcommand{\Zb}{\overline{Z}}
\newcommand{\Urm}{{\mathrm U}}
\newcommand{\Srm}{{\mathrm S}}
\newcommand{\SO}{\mathrm{SO}}
\newcommand{\Sp}{\mathrm{Sp}}
\newcommand{\SU}{\mathrm{SU}}
\newcommand{\U}{\mathrm{U}}
\newcommand{\be}{\begin{equation}}
\newcommand{\ee}{\end{equation}}
\newcommand{\Comment}[1]{{}}
\newcommand{\tQ}{\tilde{Q}}
\newcommand{\tq}{{\tilde{q}}}
\newcommand{\trho}{\tilde{\rho}}
\newcommand{\tphi}{\tilde{\phi}}
\newcommand{\Qcal}{\mathcal{Q}}
\newcommand{\tmu}{\tilde{\mu}}
\newcommand{\dbar}{\bar{\partial}}
\newcommand{\p}{\partial}
\newcommand{\eg}{{\it e.g.\;}}
\newcommand{\ie}{{\it i.e.\;}}
\newcommand{\Tr}{\mathrm{Tr}}
\newcommand{\twistor}{\Cset \mathrm{P}^{3}}
\newcommand{\note}[2]{{\footnotesize [{\sc #1}}---{\footnotesize   #2]}}
\newcommand{\CL}{\mathcal{L}}
\newcommand{\CJ}{\mathcal{J}}
\newcommand{\CA}{\mathcal{A}}
\newcommand{\CH}{\mathcal{H}}
\newcommand{\CD}{\mathcal{D}}
\newcommand{\CE}{\mathcal{E}}
\newcommand{\CQ}{\mathcal{Q}}
\newcommand{\CB}{\mathcal{B}}
\newcommand{\CC}{\mathcal{C}}
\newcommand{\CO}{\mathcal{O}}
\newcommand{\CT}{\mathcal{T}}
\newcommand{\CI}{\mathcal{I}}
\newcommand{\CN}{\mathcal{N}}
\newcommand{\CS}{\mathcal{S}}
\newcommand{\CM}{\mathcal{M}}

\begin{minipage}[t]{17.5cm}

\hfill EFI-16-02

\hfill YITP-15-123
\end{minipage}
\\ \\
\title{\Large {\bf A Small Deformation of a Simple Theory}} 
\author {Matthew Buican$^{1,2}$ and Takahiro Nishinaka$^3$} 
\affiliation{$^1$CRST and School of Physics and Astronomy \\ Queen Mary University of London, London E1 4NS, UK\\ \smallskip$^2$Enrico Fermi Institute and Department of Physics, \\ The University of Chicago, Chicago, IL 60637, USA\\ \smallskip$^3$Yukawa Institute for Theoretical Physics\\ Kyoto University, Kyoto 606-8502, Japan}

\begin{abstract}
\noindent
We study an interesting relevant deformation of the simplest interacting $\CN=2$ SCFT---the original Argyres-Douglas (AD) theory. We argue that, although this deformation is not strictly speaking Banks-Zaks like (certain operator dimensions change macroscopically), there are senses in which it constitutes a mild deformation of the parent AD theory: the exact change in the $a$ anomaly is small and is essentially saturated at one loop. Moreover, contributions from IR operators that have a simple description in the UV theory reproduce a particular limit of the IR index to a remarkably high order. These results lead us to conclude that the IR theory is an interacting $\CN=1$ SCFT with particularly small $a$ and $c$ central charges and that this theory sheds some interesting light on the spectrum of its AD parent.
\end{abstract}
\maketitle

\parskip 7pt
\section*{Introduction}
Argyres-Douglas (AD) theories \cite{Argyres:1995jj,Argyres:1995xn,Eguchi:1996vu} have traditionally been thought of as relatively mysterious superconformal field theories (SCFTs). One reason for this view is the way they were initially constructed as special points in the moduli space of $\CN=2$ gauge theories where mutually non-local BPS states become simultaneously massless. This construction makes it clear that AD theories lack a local and Lorentz-invariant Lagrangian description (although they are perfectly local SCFTs). Moreover, a simple study of their chiral spectrum via the Seiberg-Witten curve reveals that they are also strongly interacting. Finally, theorems in superconformal representation theory \cite{Dolan:2002zh, Papadodimas:2009eu} guarantee that they cannot be reached by $\CN=2$-preserving conformal deformations of free theories \cite{Buican:2014hfa}.

On the other hand, there is strong evidence that AD theories are particularly simple quantum field theories (QFTs): their conformal anomalies scale linearly with the dimensions of their Coulomb branches (i.e., their ranks) \cite{Shapere:2008zf, Xie:2013jc}, and, even more interestingly, their superconformal indices (and hence their spectra) take a particularly simple form \cite{Buican:2015ina, Buican:2015hsa,Cordova:2015nma,Buican:2015tda,Song:2015wta,Cecotti:2015lab}.

This simplicity manifests itself in many ways. For example, even though the Schur limit of the index does not receive direct \lq\lq single letter" contributions from $\CN=2$ chiral operators whose vevs parameterize the Coulomb branch, the AD Schur index still \lq\lq non-perturbatively" encodes the spectrum of these operators in its pole structure \cite{Buican:2015hsa} (and so we can think of these theories as dominated by the small number of degrees of freedom parameterizing the Coulomb branch). 

One of the consequences of this paper will be to see how to make contributions from $\CN=2$ chiral operators more manifest by performing certain small deformations of the parent AD theory. The price we will pay for making these operators more visible is that we will have to break $\CN=2\to\CN=1$.

Another particularly striking fact about AD theories is that the simplest AD theory---the so-called $(A_1, A_2)$ theory---saturates a universal lower bound for the $c$ central charge of a unitary interacting $\CN=2$ SCFT \cite{Liendo:2015ofa}. Moreover, the $(A_1, A_2)$ theory has the smallest known value of $a$ for an interacting $\CN=2$ theory (recall that $a$ and $c$ are both very similar in $\CN=2$ theories \cite{Hofman:2008ar, Shapere:2008zf}). As a result, one can think of it as the simplest member of the simplest class of $\CN=2$ SCFTs \footnote{This theory was the first AD theory ever discovered. It was found by studying the singular point of the Coulomb branch of the pure $SU(3)$ theory \cite{Argyres:1995jj}. It can also be reached by a flow from the $SU(2)$ theory with one fundamental flavor \cite{Argyres:1995xn}. There may be other theories yet to be discovered with the same value of $c$, but the $(A_1, A_2)$ theory is the only known theory to saturate this bound.}.

Therefore, it is particularly interesting to study deformations of this theory, since RG intuition tells us that the resulting IR theory should be even simpler. However, the above discussion suggests that we should, at best, only find a free theory if we deform the $(A_1, A_2)$ SCFT while preserving $\CN=2$. Indeed, this is the case \cite{Xie:2013jc, Buican:2014qla}. On the other hand, we can find a more interesting IR theory if we deform the UV SCFT in such a way as to break $\CN=2\to\CN=1$ via
\begin{equation}\label{OOdef}
\delta W=\lambda\CO^2~,
\end{equation}
where $\CO$ is the dimension $6/5$ chiral primary of the $(A_1, A_2)$ theory. One reason to study \eqref{OOdef} is that, among the available $\CN=2\to\CN=1$ breaking deformations, it is the lowest-dimensional deformation that gives rise to a stable vacuum with an interacting theory at long distance (another deformation has been recently studied in \cite{Bolognesi:2015wta}) \footnote{We will not dwell on this point here, but we will simply note that the stability of the vacuum can be argued using a general spurion analysis morally similar to one used in \cite{Gaiotto:2013sma}.}.

In what follows, we will analyze the IR theory, $\CT$, resulting from the deformation in \eqref{OOdef}. In particular, we will see that
\begin{itemize}
\item[{$\bullet$}] $\CT$ is interacting.
\item[{$\bullet$}] In the IR chiral ring, we have
\begin{equation}\label{chirRingint}
\CO^2=\CO\cdot\CO_{\alpha}=0~,
\end{equation}
where $\CO_{\alpha}$ is a spin one half $\CN=1$ chiral primary that is related to $\CO$ by $\CN=2$ SUSY in the UV.
\item[{$\bullet$}] There is strong (although not entirely conclusive) evidence that the RG flow $(A_1, A_2)\rightarrow\CT$ does not have any accidental symmetries.
\item[{$\bullet$}] The central charges of $\CT$ are (modulo the caveat in the previous point)
\begin{eqnarray}\label{acIRintro}
a_{\CT}={263\over768}~, \ \ \ c_{\CT}={271\over768}~,
\end{eqnarray}
where we have chosen a normalization in which a free chiral superfield has $c=1/24$.
\end{itemize} 

In addition, we will use the superconformal index to argue that in the IR theory
\begin{itemize}
\item[{$\bullet$}] There is a semi-short multiplet with a spin half primary, $J_{\alpha}$, of dimension $D(J_{\alpha})={11\over4}$ satisfying
\begin{equation}\label{semishort}
D^{\alpha}J_{\alpha}=0~.
\end{equation}
\end{itemize}
Using the index, we will also find evidence that $\CO$, $\CO_{\alpha}$, and a third chiral primary, $\CO'$ (that is also related to the other two by $\CN=2$ SUSY in the UV), exist as chiral operators in $\CT$. These operators have the following properties
\begin{itemize}
\item[{$\bullet$}] The dimensions of $\CO$, $\CO_{\alpha}$, and $\CO'$ in the IR are $D(\CO)={3\over2}$, $D(\CO_{\alpha})={7\over4}$, and $D(\CO')=2$ respectively.
\item[{$\bullet$}] Modulo the caveat involving accidental symmetries, $\CT$ has no flavor symmetries and so $\CO$, $\CO_{\alpha}$, and $\CO'$ are flavor singlets.
\end{itemize}

The apparent existence of a flavor singlet chiral primary, $\CO$, satisfying $\CO^2=0$ and having a scaling dimension that is within $5\%$ of the extrapolated dimension for the flavor singlet chiral primary $\phi$ operator in \cite{Poland:2015mta} that also satisfies $\phi^2=0$ begs the question of whether $\CT$ is the minimal $\CN=1$ SCFT discussed in \cite{Poland:2015mta} and if $\CO=\phi$. While these points give some reason to suspect this identification of theories might be correct, the value of the extrapolated $c$ central charge in \cite{Poland:2015mta} is roughly a factor of three smaller than the central charge in \eqref{acIRintro}. Therefore, even though $\CT$ has tantalizing similarities to the theory discussed in \cite{Poland:2015mta}, we cannot definitively conclude this is the case.

On the other hand, our study of this SCFT will shed new light on the $(A_1, A_2)$ theory and on certain aspects of $\CN=1$ dynamics. Moreover, the values of the central charges in \eqref{acIRintro} are particularly small for an interacting $\CN=1$ SCFT in four dimensions \footnote{We do not claim that the values in \eqref{acIRintro} are the smallest allowed central charges for an interacting SCFT in four dimensions. For example, \cite{Giacomelli:2014rna} argues that a particular linear deformation of the $(A_1, A_4)$ SCFT leads to an interacting IR theory with even smaller central charge (see also the discussion in \cite{Xie}). It would be interesting to apply some of our techniques to study this case as well.}. Therefore, $\CT$ clearly deserves to be studied in its own right.

The plan of the paper is as follows. In the next section, we will construct our theory and establish \eqref{chirRingint} and \eqref{acIRintro}. In the following section we will use recent insights into the superconformal indices of AD theories to argue that $\CT$ is interacting. We will then discuss constraints on accidental symmetries under the deformation in \eqref{OOdef}. In the following section, we use the index to find evidence for the existence of the primaries $\CO$, $\CO_{\alpha}$, $\CO'$, and $J_{\alpha}$ in the IR. Finally, we conclude with a discussion of the implications of our results for the conformal bootstrap and the discussion of \cite{Poland:2015mta}.

\smallskip
\noindent
{\it {\bf Note added:} While our work was being finalized, \cite{Xie} appeared. This paper has overlap with our section I---in particular with the central charge calculation and our description of the chiral ring (our calculations agree with theirs). On the other hand, our two papers are largely complementary. Indeed, \cite{Xie} motivates additional conjectures regarding $\CN=2$-perserving chiral ring relations (their equation (11)) that are compatible with our results, while our paper discusses aspects of non-chiral operators, the superconformal index, accidental symmetries, and absence of free fields.}

\section{The minimal $\CN=1$ deformation}
\label{properties}
We will make one assumption in studying the deformation \eqref{OOdef}: there are no accidental flavor symmetries along the corresponding RG flow. In section \ref{accsymm}, we will give some justifications for this assumption, but we will simply accept it for now.

From this starting point, we can compute $a_{\CT}$ and $c_{\CT}$ using 't Hooft anomaly matching and the known anomalies for the $(A_1, A_2)$ theory \cite{Aharony:2007dj,Shapere:2008zf}
\begin{equation}\label{acA1A2}
a_{(A_1, A_2)}={43\over120}~, \ \ \ c_{(A_1, A_2)}={11\over30}~.
\end{equation}
Indeed, since the $(A_1, A_2)$ theory has no $\CN=2$ flavor symmetries (as we will see momentarily it has a $U(1)$ flavor symmetry when regarded as an $\CN=1$ theory), there is a unique preserved $R$ symmetry along the RG flow
\begin{equation}\label{RGflowR}
\hat R=-2(r-{7\over12}J)={1\over6}(-5r+7R)~,
\end{equation}
where $r$ is the overall $U(1)_R$ superconformal $R$ charge in the $\CN=2$ superconformal algebra, $R$ is the $SU(2)_R$ Cartan, and $J$ is the $\CN=1$ flavor symmetry
\begin{equation}\label{defJ}
J=r+R~.
\end{equation}
We adopt the conventions $r(Q^2_{\alpha})=-R(Q^2_{\alpha})=1/2$ so that the $J(Q^2_{\alpha})=0$ (we are integrating the deformation \eqref{OOdef} over the half of superspace corresponding to $\tilde Q_{2\dot\alpha}$ and $Q^2_{\alpha}$ \footnote{We therefore define $\CO_{\alpha}=\left[Q^1_{\alpha}, \CO\right]$ and $\CO'=\left[(Q^1)^2,\CO\right]$.}). With these normalizations, we have
\begin{equation}
r(\CO)=J(\CO)=-{6\over5}~, \ \ \ R(\CO)=0~,
\end{equation}
from which \eqref{RGflowR} follows.

Next, using the well-known fact that the 't Hooft anomalies are given by \cite{Shapere:2008zf} (our conventions are $r_{\rm}=-{1\over2}R_{\CN=2}$, where $R_{\CN=2}$ is defined in \cite{Shapere:2008zf})
\begin{equation}\label{acthooft}
\CA(r^3)=-6(a-c)~, \ \ \CA(rR^2)=-(2a-c)~,\ \ \ \CA(r)=-24(a-c)~,
\end{equation}
with all other $R$ current anomalies vanishing. For the preserved $R$ symmetry, we have
\begin{equation}\label{preservedTr}
\CA(\hat R)=-{1\over6}~, \ \ \ \CA(\hat R^3)={251\over216}~.
\end{equation}
From this discussion, we conclude that
\begin{eqnarray}\label{acIR}
a_{\CT}&=&{5\over256}\left(52a_{(A_1, A_2)}-3c_{(A_1, A_2)}\right)={263\over768}~, \nonumber\\ \ \ \ c_{\CT}&=&{5\over256}\left(-12a_{(A_1, A_2)}+61c_{(A_1, A_2)}\right)={271\over768}~.\ \ \ 
\end{eqnarray}
These equations are the promised result \eqref{acIRintro} from the introduction. Note that \eqref{acIR} is compatible with the bounds in \cite{Hofman:2008ar}.

Moreover, in the IR, $\CO^2$ is a descendant since it breaks the $J$ symmetry in \eqref{defJ}
\begin{equation}\label{EOM}
\tilde D^2J\sim\lambda\CO^2~.
\end{equation}
From this equation of motion, we see that, as promised in \eqref{chirRingint}, $\CO^2$ is trivial in the IR chiral ring, i.e., $\CO^2=0$.

In fact, we can get more information by studying the $\CN=2$ supercurrent multiplet (see, e.g., \cite{Antoniadis:2010nj}). This multiplet contains an $\CN=1$ submultiplet, $J_{\alpha}$, with a primary of dimension $5/2$ and the (broken) second supersymmetry current. In the absence of supersymmetry breaking, it satisfies $\tilde D^2J_{\alpha}=0$. However, in the presence of the SUSY breaking deformation \eqref{OOdef}, we find
\begin{equation}\label{EOMII}
\tilde D^2J_{\alpha}\sim\lambda\CO\cdot\CO_{\alpha}~.
\end{equation}
Therefore, as promised in \eqref{chirRingint}, $\CO\cdot\CO_{\alpha}$ is trivial in the IR chiral ring.

\section{$\CT$ is Interacting}
\label{iprop}
To gain further insight into the IR theory, $\CT$, we will find it useful to study the superconformal index of the $(A_1, A_2)$ theory. Recall that in an $\CN=2$ theory, the index can be defined as follows
\begin{equation}\label{fullN2index}
\CI(p,q,t)={\rm Tr}(-1)^Ft^{R+r}p^{j_2-j_1-r}q^{j_2+j_1-r}e^{-\beta\Delta}~,
\end{equation}
where $R$, $r$, and $j_{1,2}$ are the $SU(2)_R$ Cartan, the overall superconformal $U(1)_R$ generator, and the two Cartans of the rotation group respectively. Note that the contributions to the trace come from states that are annihilated by $\tilde Q_{2\dot-}$ (i.e., states that have $\Delta={1\over2}\left\{\tilde Q_{2\dot-},\tilde Q_{2\dot-}^{\dagger}\right\}={1\over2}(E-2j_2-2R+r)=0$), that the fugacities $p$, $q$, $t$ satisfy $|p|,|q|,|t|, |pq/t|<1$, and that the corresponding charges also commute with $\tilde Q_{2\dot-}$ (for simplicity, we have dropped the dependence on potential flavor fugacities). While the full indices of AD theories are not presently known, results are known for various special limits \cite{Buican:2015ina, Buican:2015hsa, Cordova:2015nma, Buican:2015tda, Song:2015wta,Cecotti:2015lab}. In particular, we will find the Schur limit of the $(A_1, A_2)$ index \cite{Cordova:2015nma} to be useful below.

This special limit is defined by taking $t=q$ in \eqref{fullN2index}. As a result, all contributing states are annihilated by both $\tilde Q_{2\dot-}$ and $Q_-^1$. Using the fact that $\left\{Q^1_-,Q^{1\dagger}_-\right\}={1\over2}(E-2j_1-2R-r)$ and recalling that contributions to the index satisfy $E=2j_2+2R-r$, we see that for the contributing states in \eqref{fullN2index}, $\left\{Q^1_-,Q^{1\dagger}_-\right\}=j_2-j_1-r$. Therefore, we conclude that the Schur index is independent of $p$.

Using this freedom, we can take $p=q^{5\over7}$ and find
\begin{equation}\label{Schur}
\CI_S(q)=\CI(q^{5\over7},q,q)={\rm Tr}(-1)^Fq^{{1\over7}(12j_2+2j_1+6\hat R)}e^{-\beta\Delta}~.
\end{equation}
In particular, we see that this index is explicitly preserved when we turn on our $\CN=2\to\CN=1$ breaking deformation in \eqref{OOdef} \footnote{Recall that we are preserving the supercharge $\tilde Q_{2\dot-}$ that the index is taken with respect to since our Grassman integration measure for $\int d^2\theta\delta W+{\rm h.c.}$ is over the half of superspace corresponding to $\tilde Q_{2\dot\alpha}$ and $Q^2_{\alpha}$.}. Moreover, from \cite{Cordova:2015nma} we know that
\begin{equation}\label{SchurA1A2}
\CI_{S(A_1, A_2)}(q)=1+\sum_{\ell=1}^{\infty}{q^{\ell(\ell+1)}\over\prod_{k=1}^{\ell}(1-q^k)}=1+q^2+\cdots~,
\end{equation}
where the RHS is the Rogers-Ramanujan $H$ function.

After we turn on our relevant deformation, we should think of the index as corresponding (up to a pre-factor) to a twisted partition function for the massive theory on $S^1\times S^3$. This partition function does not depend on the RG scale. In the deep IR, after flowing to our SCFT $\CT$, we can often interpret the resulting partition function as an index for the IR theory that counts states annihilated by $\tilde Q_{2\dot-}$ (i.e., those states satisfying $\Delta_{IR}={1\over2}\left\{\tilde Q_{2\dot-},\tilde Q_{2\dot-}^{\dagger}\right\}=E-2j_2-{3\over2}\tilde R=0$, where $\tilde R$ is the IR $\CN=1$ superconformal $R$ symmetry) \footnote{For example, consider a free $\CN=2$ vector multiplet and imagine turning on a mass term for the chiral scalar, $\delta W=m\Phi^2$. In the IR, we find a free $\CN=1$ vector multiplet. The UV single letter Schur index is $I_S^{\rm s.l.}(q)=-{2q\over1-q}$ \cite{Gadde:2011uv}. Now, we can use the lack of $p$ dependence of the UV index to set $p=q$ so that $\CI_S(q)=\CI(q,q,q)={\rm Tr}(-1)^Fq^{2j_2+\hat R'}e^{-\beta\Delta}$, where $\hat R'=R-r$ is the preserved $R$ symmetry along the flow. Using this definition, we see that the IR single letter index is also $I_S^{\rm s.l.}(q)=-{2q\over1-q}$.}. More generally, if we start from a well-defined index in the UV (in particular, we need a discrete spectrum and finite index), and our relevant deformation leads to a stable vacuum, then the partition function should interpolate to the IR index or to a suitable continuation of the IR index.

As we will now see, we can use this logic to rule out the possibility that $\CT$ is a collection of free fields. In particular, the $a$-theorem \cite{Komargodski:2011vj} guarantees that the IR SCFT can at most consist of {\bf(a)} seventeen free chiral multiplets and no vector multiplets or {\bf(b)} at most eight free chiral multiplets and an abelian vector multiplet. Neither of these possibilities can reproduce \eqref{SchurA1A2}.

To understand this claim, let us study case {\bf(a)} first. We have a collection of free chiral multiplets, $\phi_i$, with $\hat R$ charges $\hat R_i$. Recall that in our conventions, contributions to the IR index come from operators that satisfy $\Delta_{\rm IR} = E-2j_2-{3\over2}\tilde R=0$, where $\tilde R$ is the free superconformal $R$ symmetry (i.e., the symmetry that assigns the $\phi_i$ charge $2/3$). These contributions can only come from states built out of bosonic chiral primaries, $\phi_i$, anti-chiral fermions, $\tilde\psi_{i\dot+}$, and their derivatives.

If we think in terms of the $S^1\times S^3$ partition function, then it is natural to consider theories with $\hat R_i\in(0,2)$ since in this case the curved space potential is bounded from below (moreover, the index is absolutely convergent)---see \cite{Gerchkovitz:2013zra, Assel:2015nca} for further discussions. However, it is easy to see that such a theory cannot reproduce \eqref{SchurA1A2} in the IR. 

Indeed, we have that 
\begin{equation}\label{IRUVindex}
\CI_{IR}(q)=\prod_{i=1}^N\prod_{m,\ell\ge0}{1-q^{{6\over7}(2-\hat R_i)+{5\over7}m+\ell}\over1-q^{{6\over7}\hat R_i+{5\over7}m+\ell}}~,
\end{equation}
where $N\le17$. We have some boson(s), $\phi_a$, of lowest $R$ charge, $\hat R_{\rm min}\in(0,2)$. In order to match \eqref{SchurA1A2}, we see that the zero-derivative single-letter contributions of the $\phi_a$ must be cancelled by fermionic contributions from some $\tilde\Psi_a$ (since the bosonic contributions appear at order less than $\CO(q^2)$ in the index). If the $\tilde\Psi_a$ are composites (in the $\tilde\psi_{i\dot+}$, $\phi_i$, and derivatives), then there are index contributions of lower order  than the index contributions of the $\phi_a$, and these contributions cannot be cancelled, which is in contradiction with \eqref{SchurA1A2}. On the other hand, if $\tilde\Psi_a=\tilde\psi_{i_a\dot+}$, then we have an exact pairing $\phi_a\oplus\tilde\psi_{i_a}$ and $\phi_{i_a}\oplus\tilde\psi_i$. Therefore, the corresponding contributions to the index cancel pairwise. We can then proceed iteratively through the remaining degrees of freedom and find that the IR index is unity. In particular, we see that \eqref{IRUVindex} cannot match the UV index.

More generally, we can ask if $\CT$ can be free if we allow some $\hat R_i\not\in(0,2)$. In this case we can try to define the index by a suitable continuation. More precisely, we start from the index of free chiral superfields
\begin{equation}\label{IRfull}
\CI_{IR}(q)=\prod_{i=1}^N\prod_{m,\ell\ge0}{1-q^{{6\over7}(2-\tilde R_i)+{5\over7}m+\ell}u_i^{-1}\over1-q^{{6\over7}\tilde R_i+{5\over7}m+\ell}u_i}~,
\end{equation}
with fugacities $u_i$ for the symmetries that act on $\phi_i$ with charge one and leave the other primaries invariant. In particular, taking $u_i\to q^{{6\over7}\alpha_i}$ so that $\hat R_i=\tilde R_i+\alpha_i$, we can obtain 
\begin{equation}\label{acont}
\tilde\CI_{IR}(q)=\prod_{i=1}^N\prod_{m,\ell\ge0}{1-q^{{6\over7}(2-\hat R_i)+{5\over7}m+\ell}\over1-q^{{6\over7}\hat R_i+{5\over7}m+\ell}}~,
\end{equation}
with some of the $\hat R_i\not\in(0,2)$. In \eqref{acont}, we have added a tilde over $\CI_{IR}$ to remind ourselves that this is a  continued expression for the index. This continuation is well-defined and non-vanishing so long as $\hat R_i\ne-{5\over6}m_i-{7\over6}\ell_i$ and $\hat R_i\ne 2+{5\over6}m_i'+{7\over6}\ell_i'$ for all non-negative integers $m_i, m_i', \ell_i, \ell_i'$.

Now, we can rewrite \eqref{acont} as follows
\begin{eqnarray}\label{IRfactor}
\tilde\CI_{IR}&=&\prod_{a=1}^{N_-}\prod_{m,\ell\ge0}\left(1\over1-q^{{6\over7}\hat R_a+{5\over7}m+\ell}\right)\cdot\prod_{A=1}^{N_+}\prod_{m,\ell\ge0}\nonumber\\ &&\left(1-q^{{6\over7}(2-\hat R_A)+{5\over7}m+\ell}\right)\cdot \tilde\CI_{IR}'~,
\end{eqnarray}
where the first factor contains the contributions of the bosons with $\hat R_a<0$, the second factor contains the contributions of the fermions coming from superfields conjugate to chiral multiplets with $\hat R_A>2$, and $\tilde\CI_{IR}'$ contains contributions from the remaining degrees of freedom. Moreover, we can rewrite the products over the $\hat R_a$ and $\hat R_A$ in \eqref{IRfactor} as follows
\begin{eqnarray}\label{IRfactorII}
&&\prod_{a=1}^{N_-}\prod_{m,\ell\ge0}\left(1\over1-q^{{6\over7}\hat R_a+{5\over7}m+\ell}\right)\nonumber\cdot\prod_{A=1}^{N_+}\prod_{m,\ell\ge0}\nonumber\\&&\left(1-q^{{6\over7}(2-\hat R_A)+{5\over7}m_A+\ell_A}\right)\nonumber\\ &&=\prod_{a=1}^{N_-}\prod_{m_a=0}^{M_{a}}\prod_{\ell_a=0}^{L_a(m_a)}\left(1\over 1-q^{{6\over7}\hat R_a+{5\over7}m_a+\ell_a}\right)\cdot\\ &&\ \ \cdot\prod_{A=1}^{N_+}\prod_{m_A=0}^{M_A}\prod_{\ell_a=0}^{L_A(m_A)}\left(1-q^{{6\over7}(2-\hat R_A)+{5\over7}m_A+\ell_A}\right)\cdots~,\nonumber
\end{eqnarray}
where we have separated contributions with ${6\over7}\hat R_a+{5\over7}m_a+\ell_a<0$ in the product over the $\hat R_a$ and fermionic contributions with ${6\over7}(2-\hat R_A)+{5\over7}m_A+\ell_A<0$ in the product over the $\hat R_A$ (all other terms, with sufficiently many derivatives so that they give rise to contributions with positive powers of $q$, appear in the ellipsis).

Note that none of the contributions to the IR index can come from contributions appearing explicitly in \eqref{IRfactorII}. Indeed, if this statement did not hold, then, by acting with sufficiently many derivatives, we would get contributions that render the IR index vanishing or ill-defined. Therefore, the bosonic and fermionic factors with the most negative $q$ exponents in \eqref{IRfactorII} must cancel. Such terms necessarily come from contributions of the $\phi_a$ with the most negative $\hat R_a<0$ and the $\tilde\psi_{A\dot+}$ with the most negative $2-\hat R_A$. In particular, we see that $\hat R_a=2-\hat R_{A}$ and that therefore the $\phi_a$ pair up with the $\tilde\psi_{A\dot+}$ and cancel in the index (similarly, the $\tilde\psi_{a\dot+}$ pair up with the $\phi_A$ and cancel). We can proceed this way iteratively through all the degrees of freedom having $\hat R_a<0$ and $\hat R_A>2$. In particular, we are back to the previous case with $\hat R_i\in(0,2)$, and so we see that the IR theory cannot consist solely of free chiral superfields.

Let us now consider case {\bf(b)}. This time, the IR index takes the form (note that here we only explicitly consider the case of uncharged matter; the case of charged matter can also be ruled out by similar means)
\begin{equation}\label{IRUVindexb}
\CI_{IR}(q)=\prod_{k\ge1}(1-q^k)(1-q^{{5\over7}k})\prod_{i=1}^N\prod_{m,\ell\ge0}{1-q^{{6\over7}(2-\hat R_i)+{5\over7}m+\ell}\over1-q^{{6\over7}\hat R_i+{5\over7}m+\ell}}~,
\end{equation}
where $N\le8$. We again should insist on $\hat R_i\ne-{5\over6}m_i-{7\over6}\ell_i$ and $\hat R_i\ne 2+{5\over6}m_i'+{7\over6}\ell_i'$ for all non-negative integers $m_i, m_i', \ell_i, \ell_i'$. Just as in the previous case, we can rule out $\hat R_i\not\in(0,2)$ since the contributions of the vector multiplet (the first product in \eqref{IRUVindexb}) cannot cancel such contributions. To see that a theory with all the $\hat R_i\in(0,2)$ cannot reproduce \eqref{SchurA1A2}, note that any $\phi_a$ with $\hat R_a\in(0,{5\over6})$ must have its index contribution cancelled by a $\tilde\psi_{A\dot+}$ with $\hat R_A\in({7\over6},2)$. In particular, $\hat R_a=2-\hat R_A$. 

Therefore, the only non-vanishing contributions to the index and to anomalies from matter fields must come from degrees of freedom with $\hat R_i\in({5\over6},2)$. Now, in order to match the 't Hooft anomaly for $\CA(\hat R^3)$ in \eqref{preservedTr}, these matter fields must make a contribution
\begin{equation}
\CA(\hat R^3)_{IR}^{\rm matter}={35\over216}~,
\end{equation}
since the gaugino makes a contribution of $+1$. In particular, there must be at least one matter field, $\phi_{i'}$, with $\hat R_{i'}>1$. As a result, we find that the matter contribution to the linear 't Hooft anomaly is
\begin{equation}
\CA(\hat R)_{IR}^{\rm matter}>-{7\over6}~.
\end{equation}
Therefore, after including the IR contribution of the gaugino (again $+1$), we find that the UV and IR linear $\hat R$ anomalies cannot match. We conclude that the IR theory is interacting.

\section{Constraints on Accidental Symmetries}
\label{accsymm}
One reason to be skeptical about the appearance of accidental symmetries is that there are no apparent unitarity bound violations. For example, if $\CO$, $\CO_{\alpha}$, and $\CO'$ exist in the IR chiral ring their dimensions are above the relevant unitarity bounds (see the introduction). Moreover, none of the UV degrees of freedom in the Schur sector have any apparent unitarity bound violations in the IR. For example, we will argue that the  non-chiral IR $\CN=1$ operator, $J_{\alpha}$, which descends from the UV $\CN=2$ stress tensor multiplet, has dimension $11/4>3/2$.

Another reason to doubt the existence of accidental symmetries in the IR comes from the fact that the one-loop change in $a$ is very close to the value we compute using $\hat R$. In particular,
\begin{eqnarray}\label{a1loop}
\delta a_{\rm 1-loop}&=&-2\pi^4\int_{0}^{\lambda_*}d\lambda\cdot\beta={1\over8}\tau_U={11\over640}\sim{61\over3840}\nonumber\\&=&a_{(A_1,A_2)}-a_{\CT}~,
\end{eqnarray}
where $\beta={3\over25}\lambda(-5+12\pi^4\tau_U^{-1}\lambda^2+\cdots)$, we have taken $\CO^2$ to have unit normalization in the UV, $a_{\CT}$ is defined in \eqref{acIR}, and \footnote{See \cite{Buican:2011ty,Buican:2013ica} for further discussions of $\tau_U$. Note that our normalization for $a$ differs from the one in \cite{Buican:2013ica}, and this explains the different numerical factor in \eqref{a1loop}.}
\begin{eqnarray}\label{tauU}
\tau_U=-{27\over4}\CA(\tilde R_{UV}(\tilde R_{UV}-\hat R)^2)={11\over80}~.
\end{eqnarray} 
Note that therefore we have $\delta a_{\rm 1-loop}\sim a_{(A_1, A_2)}-a_{\CT}\ll a_{(A_1, A_2)}$. As a result, we see that the one-loop fixed point seems to yield a consistent and surprisingly good approximation of $\CT$ (note that the coupling for the unit normalized deformation flows to a one-loop value of $\lambda_*\sim.02$) \footnote{This statement holds in spite of the fact that certain operator dimensions change macroscopically.}.

Before concluding this section, we should note that it is possible for accidental symmetries to occur in conformal perturbation theory \cite{Chester:2015qca}. However, the three-dimensional models of \cite{Chester:2015qca} do not satisfy the analogous condition to the one described in \eqref{a1loop}, i.e., they do not satisfy $\delta F_{\rm 1-loop}\sim F_{UV}-F_{IR}$ \footnote{To understand this statement, recall that the three-dimensional $\CN=2$ theories in \cite{Chester:2015qca} involve $N+1$ chiral superfields: $X$ and $Z_i$ (with $i=1, \cdots, N$). Accidental symmetries arise if we deform the interacting UV fixed point with $W={g_2\over6}X^3$ and $U(N)\times Z_3$ symmetry by the relevant operator $\delta W={g_1\over2}X\sum_i(Z_i)^2$. This deformation breaks the global symmetry to $O(N)\times Z_3$. In the IR, one can show that the putative fixed point with $W={g_2\over6}X^3+{g_1\over2}X\sum_i(Z_i)^2$ and all couplings non-zero does not exist for $N>2$ \cite{Chester:2015qca} (instead, $g_2=0$ and we have the symmetry enhancement $O(N)\times Z_3\to O(N)\times U(1)$). We claim that these putative flows do not satisfy the analog of \eqref{a1loop}. To see this, we consider $\beta_{g_1}=-{1\over3}g_1+{16\pi\over N}kg_1^3$ (where $k$ is the norm of the relevant deformation) near the interacting UV fixed point and ignore the back-reaction on the $g_2$ coupling. We then have that $\delta F_{\rm 1-loop}=-4\pi^3k\int_0^{g_1^*} dg_1\beta_{g_1}={\pi^2N\over9}\gg .056 N\sim F_{UV}-F_{IR}$, where $F_{UV}$ and $F_{IR}$ have been computed using localization \cite{Chester:2015qca}.}.

\section{Comments on IR Operators}\label{IRoperators}
In this section, we would like to motivate the existence of the $\CO$, $\CO_{\alpha}$, and $\CO'$ chiral primaries in the IR SCFT, $\CT$. As we will see, assuming these operators exist, we can reproduce the superconformal index to a very non-trivial order in $q$.

To that end, note that the IR single letter contributions of these operators are
\begin{eqnarray}
\CI_S^{\rm s.l.}(\CO)&=&{q^{6\over7}\over(1-q)(1-q^{5\over7})}~, \cr \CI^{\rm s.l.}_S(\CO_{\alpha})&=&-{q^{6\over7}+q^{8\over7}\over(1-q)(1-q^{5\over7})}~, \nonumber\\ \CI^{\rm s.l.}_S(\CO')&=&{q^{8\over7}\over(1-q)(1-q^{5\over7})}~.
\end{eqnarray}
As one would expect, these contributions cancel since the Coulomb branch sector does not contribute to the Schur limit of the UV index.

On the other hand, we have broken $\CN=2\to\CN=1$ by turning on \eqref{OOdef}. At leading order in the Coulomb branch sector, this breaking is encoded in the relations \eqref{chirRingint}. In the index, these relations give rise to the following single letter contributions \footnote{Note that we are not sensitive to $\CN=2$-preserving chiral ring relations of the form conjectured in (11) of \cite{Xie}. On the other hand, we will find a more complete picture of how $\CN=1$ chiral constraints combine with non-chiral operators to produce the correct IR physics.}
\begin{eqnarray}
\CI^{\rm s.l.}_S(\CO^2=0)&=&-{q^{12\over7}\over(1-q)(1-q^{5\over7})}~,\nonumber\\ \CI^{\rm s.l.}_S(\CO\cdot\CO_{\alpha}=0)&=&{q^{12\over7}+q^{2}\over(1-q)(1-q^{5\over7})}~.
\end{eqnarray}
On general grounds, we know that we must also have a short $\CN=1$ supercurrent, $J_{\alpha\dot\alpha}$, in the IR. This multiplet contributes
\begin{equation}
\CI_S^{\rm s.l.}(J_{\alpha\dot\alpha})=-{q^{17\over7}+q^{19\over7}\over(1-q)(1-q^{5\over7})}~.
\end{equation}

Assuming these are the only low-order contributions to the index, we find that
\begin{equation}
\CI_{IR}(q)=1+q^2-q^{17\over7}+\cdots,
\end{equation}
and so we see that we have reproduced the IR index up to order less than $\CO(q^{17\over7})$. In fact, we can do even better. 

Indeed, recall from (the conjugate of) \eqref{EOMII} that
\begin{equation}
D^2\tilde J_{\dot\alpha}\sim \tilde\lambda\tilde\CO\cdot\tilde\CO_{\dot\alpha}~.
\end{equation}
Therefore, we might be tempted to conclude that $\tilde J_{\dot\alpha}$ is a long multiplet. However, we can see in conformal perturbation theory that
\begin{equation}
\tilde D_{\dot\alpha}\tilde J^{\dot\alpha}=0~.
\end{equation}
In particular, at a fixed point, this is the shortening condition for a multiplet of type $\bar\CC_{-{1\over6}(0,{1\over2})}$ in the $\CN=1$ classification of \cite{Beem:2012yn}. Therefore, we see it is reasonable to believe that $\tilde J_{\dot\alpha}$ exists as a short multiplet in the IR theory with dimension
\begin{equation}
D(\tilde J_{\dot\alpha})={11\over4}~.
\end{equation}
It is then straightforward to check that the single letter index contribution of this multiplet is
\begin{equation}
\CI_S^{\rm s.l.}(\tilde J_{\dot\alpha})={q^{17\over7}\over(1-q)(1-q^{5\over7})}~.
\end{equation}
Taking this contribution into account, one finds
\begin{equation}\label{IRindexfinal}
\CI_{IR}(q)=1+q^2+q^3+\CO(q^4)~,
\end{equation}
and we can reproduce the Rogers-Ramanujan $H$ function to a remarkably high order just using the $\CN=2$ stress tensor multiplet and operators from the Coulomb branch sector.

Finally, It is easy to check that
\begin{itemize}
\item[{$\bullet$}] There are no additional single letter contributions from the IR $\CN=1$ multiplets that descend from the $\CN=2$ supercurrent multiplet.
\item[{$\bullet$}] The IR contributions due to operators that are annihilated in the UV by $\tilde Q_{2\dot-}$ and sit in the remaining $\CN=2$ Schur multiplets cannot arise at order smaller than $\CO(q^{29\over7})$ \footnote{An easy way to see this bound is as follows. Such states in the UV satisfy $E-R=2j_2+R-r$. The smallest $E-R$ in the Schur multiplet is for the Schur operator itself, and we define $(E-R)_{\rm min}\equiv h$. In particular, we have that any element of the multiplet annihilated by $\tilde Q_{2\dot-}$ has $R+2j_2\ge h+r$. Therefore, we see this reasoning gives a lower bound on such contributions to the IR index since they occur at order $q^N$ with $N={1\over7}(12j_2+2j_1+7R-5r)\ge{1\over7}(6h+2j_1+R+r)$. Now, since the chiral algebra is just a Virasoro algebra, the primary, $\CO$ (we suppress possible $SU(2)_R$ and Lorentz indices), has $r=0$ and $j_1(\CO)_{\rm max}=j_2(\CO)_{\rm max}\equiv j(\CO)_{\rm max}$ with $E(\CO)-R(\CO)_{\rm max}=2+2j(\CO)_{\rm max}+R(\CO)_{\rm max}=h$. As a result, we see that $2j_1(\CO)+R(\CO)+r(\CO)\ge2-h$. Now, the smallest such quantum number for any operator in the multiplet annihilated by $\tilde Q_{2\dot-}$ is $2j_1+R+r\ge-1-h$. As a result, $N\ge{1\over7}(5h-1)$ Null state relations in the $(A_1, A_2)$ chiral algebra imply that the first Schur contribution not due to the stress tensor multiplet is at $h=6$. Therefore, we find the lower bound on contributions at $\CO(q^{29\over7})$.
}. 
\end{itemize} 

\section{Connection to the Bootstrap?}
We have seen some tantalizing similarities to the minimal theory described in \cite{Poland:2015mta} (see also \cite{Bobev:2015jxa}). For one, we seem to find a flavor singlet chiral operator, $\CO$, satisfying $\CO^2=0$ with dimension $3/2$. In \cite{Poland:2015mta}, the authors found a flavor singlet chiral operator, $\phi$, satisfying $\phi^2=0$ with dimension estimated to be $\sim1.43$. On the other hand, our value of the central charge, $c$, while being one of the smallest such values we are aware of in any existing theory, is a factor of three larger than the central charge in \cite{Poland:2015mta}.  Therefore, it presumably must be the case that either:
\begin{itemize}
\item[{$\bullet$}] Our SCFT is the minimal interacting $\CN=1$ model (as indicated by the presence of some kink in the solution space to certain crossing equations). In this case, there must be additional constraints that rule out the existence of similar SCFTs with lower $c$ like the one in \cite{Poland:2015mta} (or perhaps less likely there are subtleties with bootstrap extrapolations of $c$).
\item[{$\bullet$}] Alternatively, our SCFT is not the minimal $\CN=1$ model. In this case, it does not seem likely that we can flow from $\CT$ to the minimal theory in \cite{Poland:2015mta} since there does not seem to be a relevant SUSY deformation that leads to an interacting IR theory. Another possibility is that our SCFT $\CT$ is a direct sum of the theory in \cite{Poland:2015mta} and some free fields that we are unable to see. In this case, we can flow to the theory in  \cite{Poland:2015mta} by turning on superpotential mass terms.
\end{itemize}

\section{Conclusions}
We have learned a surprising amount by studying a simple deformation of the minimal Argyres-Douglas theory. At the level of the parent $(A_1, A_2)$ theory, we have seen evidence that the full low-lying spectrum of short multiplets is likely simpler than one might expect. Indeed, we were able to reproduce \eqref{IRindexfinal} simply from the IR descendants of the $\CN=2$ Coulomb branch and stress tensor multiplets (the existence of the semi-short $J_{\alpha}$ multiplet in the IR suggests that our deformation of the parent AD theory is particularly mild). Moreover, we saw that we could trade UV index contributions from the $SU(2)_R$ current with contributions from constrained chiral operators in the IR theory. This result points to some deeper connections between the physics of chiral algebras and $\CN=2$ chiral rings upon $\CN=2\to\CN=1$ breaking that we will return to soon.

In addition, our results point to some tantalizing potential connections with the bootstrap and theories of the type described in \cite{Poland:2015mta}. Another interesting issue that has potential overlap with the bootstrap is the following. We expect deformations of the type we introduced in \eqref{OOdef} to become mass terms when the scaling dimension of $\CO$ equals one. It would be interesting to understand if there are classes of theories in which a cross-over to a mass-term-like behavior happens when the dimension of $\CO$ is $1+\epsilon$. An initial study of the $(A_1, A_{2n})$ theories for $n>1$ (as $n\to\infty$, there are Coulomb branch operators whose dimension approaches one) suggests that in this class the cross over may not occur for any $\epsilon>0$ (although we have not yet subjected these theories to all the tests described in this note).

Finally, our result that the change in the $a$ anomaly is essential saturated at one loop \eqref{a1loop} implies that there should be some surprisingly small OPE coefficients in the $(A_1, A_2)$ theory. Understanding these coefficients more quantitatively will likely lead to a deeper understanding of the symmetries (exact and approximate) of the $(A_1, A_2)$ theory and, perhaps, of its more complicated AD siblings.

\medskip
\centerline{\bf Acknowledgements}
\noindent
We would like to thank D.~Kutasov for many interesting discussions. M.~B.'s work is partially supported by the Royal Society under the grant \lq\lq New Constraints and Phenomena in Quantum Field Theory" and by the U.S. Department of Energy under grant DE-SC0009924. T.~N. is partially supported by the Yukawa Memorial Foundation.


\begin{thebibliography}{}
\bibitem{Argyres:1995jj} 
  P.~C.~Argyres and M.~R.~Douglas,
  Nucl.\ Phys.\ B {\bf 448}, 93 (1995)
  [hep-th/9505062].

\bibitem{Argyres:1995xn} 
  P.~C.~Argyres, M.~R.~Plesser, N.~Seiberg and E.~Witten,
  Nucl.\ Phys.\ B {\bf 461}, 71 (1996)
  [hep-th/9511154].

\bibitem{Eguchi:1996vu} 
  T.~Eguchi, K.~Hori, K.~Ito and S.~K.~Yang,
  Nucl.\ Phys.\ B {\bf 471}, 430 (1996)
  [hep-th/9603002].

\bibitem{Dolan:2002zh} 
  F.~A.~Dolan and H.~Osborn,
  Annals Phys.\  {\bf 307}, 41 (2003)
  [hep-th/0209056].

\bibitem{Papadodimas:2009eu} 
  K.~Papadodimas,
  JHEP {\bf 1008}, 118 (2010)
  [arXiv:0910.4963 [hep-th]].

\bibitem{Buican:2014hfa} 
  M.~Buican, S.~Giacomelli, T.~Nishinaka and C.~Papageorgakis,
  JHEP {\bf 1502}, 185 (2015)
  [arXiv:1411.6026 [hep-th]].

\bibitem{Shapere:2008zf} 
  A.~D.~Shapere and Y.~Tachikawa,
  JHEP {\bf 0809}, 109 (2008)
  [arXiv:0804.1957 [hep-th]].

\bibitem{Xie:2013jc} 
  D.~Xie and P.~Zhao,
  JHEP {\bf 1303}, 006 (2013)
  [arXiv:1301.0210].


\bibitem{Buican:2015ina} 
  M.~Buican and T.~Nishinaka,
  arXiv:1505.05884 [hep-th].

\bibitem{Buican:2015hsa} 
  M.~Buican and T.~Nishinaka,
  arXiv:1505.06205 [hep-th].

\bibitem{Cordova:2015nma} 
  C.~Cordova and S.~H.~Shao,
  arXiv:1506.00265 [hep-th].

\bibitem{Buican:2015tda} 
  M.~Buican and T.~Nishinaka,
  arXiv:1509.05402 [hep-th].

\bibitem{Song:2015wta} 
  J.~Song,
  arXiv:1509.06730 [hep-th].

\bibitem{Cecotti:2015lab} 
  S.~Cecotti, J.~Song, C.~Vafa and W.~Yan,
  arXiv:1511.01516 [hep-th].

\bibitem{Liendo:2015ofa} 
  P.~Liendo, I.~Ramirez and J.~Seo,
  arXiv:1509.00033 [hep-th].

\bibitem{Hofman:2008ar} 
  D.~M.~Hofman and J.~Maldacena,
  JHEP {\bf 0805}, 012 (2008)
  [arXiv:0803.1467 [hep-th]].

\bibitem{Buican:2014qla} 
  M.~Buican, T.~Nishinaka and C.~Papageorgakis,
  JHEP {\bf 1412}, 095 (2014)
  [arXiv:1407.2835 [hep-th]].

\bibitem{Bolognesi:2015wta} 
  S.~Bolognesi, S.~Giacomelli and K.~Konishi,
  JHEP {\bf 1508}, 131 (2015)
  [arXiv:1505.05801 [hep-th]].

\bibitem{Poland:2015mta} 
  D.~Poland and A.~Stergiou,
  arXiv:1509.06368 [hep-th].

\bibitem{Giacomelli:2014rna} 
  S.~Giacomelli,
  JHEP {\bf 1501}, 044 (2015)
  [arXiv:1409.3077 [hep-th]].

\bibitem{Xie} 
  D.~Xie and K.~Yonekura,
  [arXiv:1602.04817 [hep-th]].

  
\bibitem{Aharony:2007dj} 
  O.~Aharony and Y.~Tachikawa,
  JHEP {\bf 0801}, 037 (2008)
  [arXiv:0711.4532 [hep-th]].

\bibitem{Antoniadis:2010nj} 
  I.~Antoniadis and M.~Buican,
  JHEP {\bf 1104}, 101 (2011)
  [arXiv:1005.3012 [hep-th]].

\bibitem{Gaiotto:2013sma} 
  D.~Gaiotto, S.~Gukov and N.~Seiberg,
  JHEP {\bf 1309}, 070 (2013)
  [arXiv:1307.2578 [hep-th]].
\bibitem{Komargodski:2011vj} 
  Z.~Komargodski and A.~Schwimmer,
  JHEP {\bf 1112}, 099 (2011)
  [arXiv:1107.3987 [hep-th]].

\bibitem{Gerchkovitz:2013zra} 
  E.~Gerchkovitz,
  JHEP {\bf 1407}, 071 (2014)
  [arXiv:1311.0487 [hep-th]].

\bibitem{Assel:2015nca} 
  B.~Assel, D.~Cassani, L.~Di Pietro, Z.~Komargodski, J.~Lorenzen and D.~Martelli,
  JHEP {\bf 1507}, 043 (2015)
  [arXiv:1503.05537 [hep-th]].

\bibitem{Buican:2011ty} 
  M.~Buican,
  Phys.\ Rev.\ D {\bf 85}, 025020 (2012)
  [arXiv:1109.3279 [hep-th]].

\bibitem{Buican:2013ica} 
  M.~Buican,
  JHEP {\bf 1401}, 155 (2014)
  [arXiv:1311.1276 [hep-th]].

\bibitem{Chester:2015qca} 
  S.~M.~Chester, S.~Giombi, L.~V.~Iliesiu, I.~R.~Klebanov, S.~S.~Pufu and R.~Yacoby,
  arXiv:1507.04424 [hep-th].

\bibitem{Beem:2012yn} 
  C.~Beem and A.~Gadde,
  JHEP {\bf 1404}, 036 (2014)
  [arXiv:1212.1467 [hep-th]].
  
\bibitem{Gadde:2011uv} 
  A.~Gadde, L.~Rastelli, S.~S.~Razamat and W.~Yan,
  Commun.\ Math.\ Phys.\  {\bf 319}, 147 (2013)
  doi:10.1007/s00220-012-1607-8
  [arXiv:1110.3740 [hep-th]].

\bibitem{Bobev:2015jxa} 
  N.~Bobev, S.~El-Showk, D.~Mazac and M.~F.~Paulos,
  JHEP {\bf 1508}, 142 (2015)
  [arXiv:1503.02081 [hep-th]].


\end{thebibliography}
\end{document}